\documentclass[seceq,supplement]{ptptex}

\usepackage{graphicx}

\newcommand{\ee}{\mathrm{e}}
\newcommand{\ii}{\mathrm{i}}
\newcommand{\C}{\mathcal{C}}
\newcommand{\WW}{\mathbb{W}}
\newcommand{\HH}{\mathbb{H}}
\newcommand{\BB}{\mathcal{B}}
\newcommand{\Acal}{\mathcal{A}}  
\newcommand{\ppt}{\frac{\partial}{\partial t}}
\newcommand{\sig}{\sigma}
\newcommand{\hf}{\frac12}

\newcommand{\ssket}{|\mathrm{ss}\rangle}
\newcommand{\pss}{p^\mathrm{ss}}

\title{
Large deviations and ensembles of trajectories in stochastic models
}

\author{
Robert L. Jack$^1$
and 
Peter Sollich$^2$ 
}

\inst{
$^1$ Department of Physics, University of Bath, Bath BA2 7AY, UK.
\\
$^2$ King's College London, Department of Mathematics, London WC2R 2LS, UK. 
}

\abst{
We consider ensembles of trajectories associated with 
large deviations of time-integrated quantities in stochastic
models.  Motivated by proposals that these ensembles are
relevant for physical processes such as shearing and
glassy relaxation, we show how they can be generated
directly using auxiliary stochastic processes.  We illustrate our
results using the Glauber-Ising chain, for which biased ensembles
of trajectories can exhibit ferromagnetic ordering.  We
discuss the relation between such biased ensembles and
quantum phase transitions.
}

\begin{document}

\maketitle

\section{Introduction}

This article is concerned with ensembles of trajectories (time-realisations)
of stochastic model systems, in which we impose constraints on time-integrated
quantities.  For example, one might consider an system coupled to two particle
reservoirs over a time period $t$, and insist that the total current through the system takes a given
value~\cite{Bod-Der}.  Such ensembles underlie a range of general
results in non-equilibrium systems~\cite{Ruelle,fluct,ECM-93,JL,JL-control,
Lebowitz-Spohn,Maes-general} and have been used to investigate transport
properties in simple models and  specific non-equilibrium steady 
states~\cite{Bod-Der,Mike-Evans}. Recently, they have also been 
employed in studies of the glass transition~\cite{merolle,s-glass,juanpe-fred-jpa,
rom-paper,p2-paper}, 
where adjusting constraints on such ensembles can drive ergodic `model fluids'
into non-ergodic states that resemble `ideal glasses'.

The language of ensembles and constraints indicates that our methods will be 
related to those of equilibrium thermodynamics.  The crucial difference is
that we consider constraints on time-integrated quantities, while thermodynamics
is concerned with constraints that apply at all times in a system.  While these
definitions may sound similar, they typically lead to quite different behaviour.
To understand this difference in a qualitative way, we observe that macrostates in 
constrained thermodynamic systems may be obtained by minimising
the free energy, which corresponds to a 
minimisation of the work required to introduce the macroscopic constraints.
On the other hand, when time-integrated quantities are being constrained, one
must instead minimise a `dynamical free energy' that corresponds to
the power required to maintain the constraints, in the face of thermal fluctuations~\cite{JL-control}. 
For a constraint on a given quantity, minimising the work and the power are not equivalent: to minimise
the dissipated power, it is preferable for the system to enter states from which spontaneous relaxation to
equilibrium is very slow, even if the work required to generate such states is relatively large.
For example, we show below that in a $1d$ Ising chain, the power required to stabilise
low energy states is minimised if the system develops ferromagnetic order, while the
work required to attain the same instantaneous value of the energy is minimised by
a paramagnetic state -- after all, constraining the energy to low values corresponds to considering low temperature, which in $1d$ only ever produces paramagnetic states.

Thus, the ensembles of trajectories that we consider here are not straightforwardly related
to thermodynamic ensembles.  Nevertheless, studies of the glass transition~\cite{merolle,s-glass}
and of boundary-driven sheared systems~\cite{Mike-Evans} have proposed that ensembles
of trajectories generated in this way are relevant for the results of physical experiments.
Establishing a connection between constrained ensembles of trajectories and
experimental systems is a complicated task.
For example, the most natural representation of the constrained 
ensemble does not have a Markov form for the transition rates.
Further, causality may be broken, in the sense that applying a perturbation
at time $t_\mathrm{w}$ may incur a response at times $t<t_\mathrm{w}$.
Here, we describe some general features of such constrained
ensembles of trajectories, aiming to understand what
physical protocol might lead to the same behaviour as a
constraint on a time-averaged quantity.

Our results are organised into two sections.
In Section~\ref{sec:general}, we review some general 
aspects of the large deviation formalism that we will use,
and we define an `auxiliary model' that is a Markov stochastic process whose
steady state trajectories are close to those of the constrained 
ensembles (in a sense discussed below).  A similar result
was described by Evans~\cite{Mike-Evans}: we show how the rules
discussed in that work can be derived through a similarity
transformation between master operators for stochastic processes. For cases
where the constrained ensemble of trajectories is time-reversal invariant,
the auxiliary model can be obtained though a modification of 
the energy function for the original model.  We discuss the extent
to which these auxiliary models represent physical realisations of
the constrained ensembles described above.
Then, in Section~\ref{sec:ising},
 we investigate spontaneous breaking of ergodicity
in constrained ensembles of trajectories, for
the (one-dimensional) Glauber-Ising chain.  Despite its
$1d$ nature, constraining ensembles of trajectories in this model
may induce long-ranged order and spontaneous symmetry breaking.  
The mechanism is identical to that behind quantum phase transitions,
where real time in the stochastic model plays the role of imaginary
time in a path integral representation of the density matrix.
We discuss the interpretation of the auxiliary stochastic
model in this context.

\section{Biased ensembles of trajectories}
\label{sec:general}

\subsection{Definitions}

We consider stochastic models in continuous time.
A model is defined through a (discrete) set of configurations 
$\{\C_1,\C_2,\dots\}$ and
stochastic transition rates
$W(\C'\to \C)$.  Then, if $p(\C,t)$ is the 
probability of finding the system in configuration $\C$ at
time $t$, the master equation of the model is
\begin{equation}
\ppt p(\C,t) = -r(\C) p(\C,t) + \sum_{\C'\neq \C} W(\C'\to \C) p(\C',t)
\end{equation}
where
$r(\C)=\sum_{\C'\neq \C} W(\C\to\C')$
is the escape rate from configuration $\C$.  Writing $|P(t)\rangle
=\sum_{\C} p(\C,t) |\C\rangle$, the master equation can
be written as $\ppt|P(t)\rangle=\WW|P(t)\rangle$ where the matrix
elements of the operator $\WW$ are simply $\langle \C|\WW|\C'\rangle
=W(\C'\to\C) - \delta_{\C,\C'} r(\C)$.

We consider large deviations of time-integrated quantities of the form
%
\begin{equation}
\BB_t =  \int_0^t\!\mathrm{d}t' B(t')
\end{equation}
where $B(t)$ is an observable that depends only
the configuration of the system at time $t$.  (In 
Section~\ref{sec:ising}, we will take $B(t)$ to be simply the energy of the 
system.)  
Alternatively, one can consider large deviations 
fluxes or dynamical activities.  That is, assume that an observable $\Acal_t$
acquires contributions whenever the system changes configuration:
if the sequence of configurations in a trajectory is $\C_0,\C_1,\dots,\C_K$ then
$\Acal_t$ can be written in the form
\begin{equation}
\Acal_t = \sum_{k=0}^{K-1} \alpha(\C_k,\C_{k+1}).
\label{equ:obsA}
\end{equation}
For example, one might take $\alpha(\C_k,\C_{k+1})=1$ for all pairs of
distinct configurations, so that $\Acal_t$ counts the number of configuration changes
between time $0$ and time $t$~\cite{merolle,s-glass}. Alternatively,
one might take $\alpha(\C_k,\C_{k+1})$
to be the contribution of the stochastic transition $\C_k\to\C_{k+1}$ to a total current~\cite{Bod-Der}, 
accumulated shear~\cite{Mike-Evans},
entropy current~\cite{Lebowitz-Spohn}, or dynamical entropy~\cite{fred-jsp}.  For convenience,
we concentrate on observables of the form $\BB_t$, but 
most of our results generalise straightforwardly\cite{juanpe-fred-jpa,fred-jsp}
to time-extensive observables of the form $\Acal_t$.

Following the discussion
in the introduction, we are interested in ensembles
of trajectories where $\BB_t$ is constrained to be far
from its average value  $\langle\BB_t\rangle_0$, where
$\langle \cdot \rangle_0$ indicates an average
in the steady state of the stochastic model.  However,
it is convenient to exploit an equivalence of ensembles
in the spirit of microcanonical/canonical equivalence
in statistical mechanics: instead of a constrained ensemble,
we define a `biased ensemble'
by assigning to each trajectory $\pi$ a probability that depends
on a `biasing field' $g$:
\begin{equation}
P[\pi,g] = Z(g,t)^{-1} P[\pi,0]  
  \exp\left[ -g \BB_t \right]
\label{equ:s-ens}  
\end{equation}
where $P[\pi,0]$ is the probability
of observing trajectory $\pi$ in the (unbiased) steady state of
the stochastic model and
\begin{equation}
Z(g,t) = \left\langle 
    \exp\left[ -g \BB_t\right] 
                                       \right\rangle_0
\end{equation}
is the partition function for the new ensemble.
Our formalism and notation closely follows Ref.~\citen{juanpe-fred-jpa}, and we refer to that
paper for further technical details.  
We note that the field $g$ was denoted by $s$ in Refs.~\citen{s-glass,
juanpe-fred-jpa,p2-paper,rom-paper}, and the resulting biased ensembles
accordingly referred to as $s$-ensembles.
We prefer $g$ in this article to avoid confusion with
Ising spins $s_i$ in later sections.
%
Averages within the biased ensemble are given by
\begin{equation}
\langle O \rangle_g =  Z(g,t)^{-1} \left\langle O  
   \exp\left[ -g \BB_t \right] 
   \right\rangle_0   
\end{equation}
where $O$ may be any trajectory-dependent observable, and the average
depends implicitly on the time $t$ as well as on $g$.

From a mathematical point of view, one may identify $Z(g,t)$
as the generating function for the moments of $\BB_t$.  Physically,
we note
that the effect of $g$ is to bias the average $\langle \BB_t \rangle_g$.
Then, we appeal to equivalence of ensembles:
the ensemble defined by (\ref{equ:s-ens}) is equivalent
to an ensemble in which $\BB_t$ is constrained to take the
value $\langle \BB_t \rangle_g$, where equivalence
holds in the same sense as for the canonical and microcanonical
ensembles in statistical mechanics.

Finally, we obtain 
the `dynamical free energy' of the biased ensemble.
Let $\pss(\C)$ be the probability of observing
configuration $\C$ in the (unbiased) steady state of the original model.
Then, one may write~\cite{fred-jsp}
\begin{equation}
Z(g,t)=\langle e| \ee^{\WW(g)t} \ssket
\label{equ:zmat}
\end{equation}
 where $\langle e|=\sum_\C \langle \C|$
is a projection state,  $\ssket=\sum_\C \pss(\C) |\C\rangle$,
and
\begin{equation}
\WW(g) = \WW - g \sum_\C |\C\rangle \langle \C| b(\C) 
\label{equ:WB}
\end{equation}
where $b(\C)$ is the value of the observable $B$ in configuration $\C$.  
Assuming for convenience that the original stochastic model is irreducible and 
has a finite state space,
one makes a spectral decomposition of $\WW(g)=\sum_i |V_i\rangle \omega_i \langle U_i|$,
and it follows from (\ref{equ:zmat})
that the dynamical free energy is~\cite{Lebowitz-Spohn,fred-jsp}
\begin{equation}
\psi(g) \equiv -\lim_{t\to\infty} t^{-1} \log Z(g,t) = - \max_i \omega_i .
\label{equ:def-psi}
\end{equation}
We note that 
while we assumed an irreducible model and a finite system, these are not necessary
for (\ref{equ:def-psi}) to hold.  However, infinite systems with spontaneously broken 
ergodicity~\cite{Gaveau-Schulman,rom-paper} or
absorbing states~\cite{absorbing} require some additional assumptions on the 
operator $\WW(g)$.

If we wish to consider the large deviations of
observables $\Acal_t$ defined as in (\ref{equ:obsA}),
the analysis is very similar:
Equs.~(\ref{equ:s-ens})-(\ref{equ:def-psi})
still hold, with $\BB_t$ simply replaced by $\Acal_t$.  However, the operator $\WW(g)$ 
has a different form in this case~\cite{fred-jsp,juanpe-fred-jpa}: its matrix elements are
\begin{equation}
\langle \C | \WW(g) | \C'\rangle = \left\{ \begin{array}{ll} 
W(\C'\to\C) \ee^{-g\alpha(\C',\C)}, \quad & \C\neq\C', \\
-r(\C), & \C=\C' . \end{array} \right.
\label{equ:WA}
\end{equation}

\subsection{Time-translational invariance, and an
auxiliary stochastic process that generates the biased ensemble}
\label{sec:tti}

\newcommand{\F}{F}
\newcommand{\f}{f}

We now turn to physical features of the biased ensemble
of (\ref{equ:s-ens}).
Averages $\langle O \rangle_0$ in the steady state of the
stochastic model are clearly time-translational invariant.
However, this invariance is broken for observables
such as $\langle O \rangle_g$.  For example, if one takes a
configuration-dependent observable $\F$, averages such
as $\langle \F(t') \rangle_g$ may depend on $t'$ since the bias
in (\ref{equ:s-ens}) breaks time-translation 
invariance~(TTI).  However, 
as long as the operator $\WW(g)$ is not at a critical
point (in a sense defined below),
we have
\begin{eqnarray}
\langle \F(t') \rangle_g = \left\{ \begin{array}{ll}
\F_\infty, & 
\tau \ll t', \ \tau \ll t-t'
\\
\F_\mathrm{f} & t'=
t\gg\tau
\end{array}\right.
\label{equ:F-ave}
\end{eqnarray}
where $\tau$ is a relaxation time into the TTI regime, discussed below.
In general, the biased ensembles are TTI
for a range of times $t'$ such that $\tau\ll t'$ and $\tau\ll t-t'$.

To understand the TTI regime in more detail, we consider
the operator $\WW(g)$.
This operator is not Hermitian, so it has separate left
and right eigenvectors associated with its largest eigenvalue,
which we denote by $\langle U|=\sum_\C \langle \C| u(\C)$ 
and $|V\rangle=\sum_\C v(\C) |\C\rangle$ respectively.  (We normalise such
that $\langle e | V\rangle=1=\langle U|V\rangle$. This is possible since the
$u(\C)$ and $v(\C)$ are non-negative, as will become clear from the
probability interpretation derived below.)
Then, for long times $t'$, we have 
\begin{equation}
\ee^{\WW(g) t'} =
 \ee^{-\psi(g)t'} \left[ |V\rangle\langle U| + \mathcal{O}(\ee^{-\Delta\omega t'})\right] ,
\label{equ:tti_proj}
\end{equation} 
where $\Delta \omega$ is the difference between the largest and second-largest
eigenvalues of $\WW(g)$.  If we restrict to irreducible stochastic models with
finite state spaces then $\Delta\omega>0$ and we identify $\tau=(\Delta\omega)^{-1}$.  
However, in the limit of large system
size, $\Delta\omega$ may vanish.  In some cases, this signifies a critical point
for the operator $\WW(g)$, while in other cases, the analysis of this section may
be generalised and the TTI regime still exists.  Examples of both cases are discussed
in Section~\ref{sec:ising}.

In any case, one has from~(\ref{equ:F-ave}) that
\begin{eqnarray}
\F_\infty &=& \lim_{t\to\infty} \langle e | \ee^{\WW(g)(t-t')} \hat{\F} \ee^{\WW(g)t'} \ssket Z(g,t)^{-1}
\\
\F_\mathrm{f} &= &\lim_{t\to\infty} \langle e | \hat\F \ee^{\WW(g)t} \ssket Z(g,t)^{-1}
\end{eqnarray}
where 
$\hat{\F} = \sum_\C |\C\rangle \langle \C| \f(\C)$, with
$\f(\C)$ being the value of observable $\F$ in configuration $\C$.
Then, as long as the corrections in (\ref{equ:tti_proj}) are small, one has
$Z(g,t)=\ee^{-\psi(g)t}\langle U|\mathrm{ss}\rangle$ and so
\begin{equation}
\F_\infty = \langle U|\hat{\F}|V\rangle = \sum_\C \f(\C) u(\C) v(\C),
\qquad \F_\mathrm{f} = \langle e|\hat{\F}|V\rangle = \sum_\C \f(\C) v(\C)
\end{equation} 
Taking then $\F$ to be simply an indicator function for configuration 
$\C$, one sees that
the probability of observing this configuration in the TTI
regime of the biased ensemble is 
\begin{equation}
p^\mathrm{TTI}(\C)=u(\C) v(\C),
\end{equation}
while the probability of observing the same configuration at the final time $t$
is simply $v(\C)$.  Indeed, as $t'$ is reduced from $t$,  $\langle F(t') \rangle_g$ decays 
exponentially from $\F_\mathrm{f}$ to $F_\infty$, on a time scale $\tau=(\Delta\omega)^{-1}$. 

We now construct an auxiliary
stochastic model whose (unbiased) trajectories 
coincide with those of the biased ensemble of (\ref{equ:s-ens}),
within the TTI regime. We will show that, in terms of the diagonal operator
$\hat U = \sum_\C |\C\rangle \langle\C| u(\C)$, the master operator of
this auxiliary model is
\begin{equation}
\WW^\mathrm{aux} = \hat{U} \WW(g) \hat{U}^{-1} + \psi(g).
\label{equ:w_aux}
\end{equation}
The off-diagonal matrix elements of this operator must then be the transition rates
between the configurations of the system. For biased ensembles 
defined as in (\ref{equ:s-ens}), these rates are
\begin{equation}
W^\mathrm{aux}(\C'\to\C)  \equiv
\langle \C | \WW^\mathrm{aux} |\C'\rangle
 = u(\C) W(\C'\to\C) \frac1{u(\C')},
 \label{equ:aux_rates}
\end{equation}
which are non-negative as they should be.
The model is also stochastic: $\sum_\C \langle \C| \WW^\mathrm{aux}
=\langle e|\WW^\mathrm{aux}=0$
which follows since $\langle U|=\langle e|\hat{U}$
is a left eigenvector of $\WW(g)$
with eigenvalue $-\psi(g)$.  Thus, the diagonal elements
of $\WW^\mathrm{aux}$ are simply 
$\langle \C | \WW^\mathrm{aux} |\C\rangle = - \sum_{\C'}
W^\mathrm{aux}(\C\to\C')$. 

Two comments are in order.  Firstly, if the bias in (\ref{equ:s-ens}) 
uses an observable $\Acal_t$ in place of $\BB_t$, the auxiliary rates are 
modified to
\begin{equation}
W^\mathrm{aux}(\C'\to\C)  \equiv
\langle \C | \WW^\mathrm{aux} |\C'\rangle
 = u(\C) W(\C'\to\C) \ee^{-g\alpha(\C',\C)} \frac1{u(\C')}.
\end{equation}
Secondly, applying this generalised result to an ensemble biased
by the total shear, one obtains the result of Evans~\cite{Mike-Evans}:
we identify our $u(\C)$ with his $\ee^{q_\C}$, where $q_\C$ 
is defined as a measure of ``the propensity [of configuration $\C$] to exhibit
flux in the future'', via $\ee^{q_\C}=
\lim_{t'\to\infty} 
\frac1{Z(t')}
\sum_{\C'} \langle \C'| \ee^{\WW(g) t'}|\C\rangle=u(\C)/\langle
U|\mathrm{ss}\rangle \propto u(\C)$.

We also note that
$|\mathrm{TTI}\rangle = \sum_\C p^\mathrm{TTI}(\C) |\C\rangle$
is a right eigenvector of $\WW^\mathrm{aux}$ with eigenvalue zero,
and we identify this as the steady state distribution 
of the auxiliary model, consistent with our assertion that
the steady state trajectories of the auxiliary model are those
of the biased ensemble of (\ref{equ:s-ens}) in the TTI regime.

To demonstrate that the auxiliary model is valid, we now need to show that
this correspondence holds for all trajectories
and not simply for the steady state distribution over configurations.
Consider, then, the probability that a given path
occurs within the TTI regime of the biased ensemble. We define the path $\pi$
by discretising time in the manner of a path integral.
That is, we state the configuration of the system
at a sequence of equally-spaced times: taking the spacing $\Delta t$
to zero then allows us to specify the path with arbitrary precision.
Let the sequence of configurations be
$\C_0,\C_1,\dots,\C_M$, where $\C_0$ occurs at
a time $t_0$ so that $\C_M$ occurs at time $t_0+M\Delta t$. To ensure
that we are in the TTI regime we take $\tau\ll t_0$ and $\tau \ll t-(t_0+M\Delta t)$.
Then, the probability of the path $\pi$ in the biased ensemble is simply
\begin{equation}
P[\pi,g] = \left[\sum_\C G_{\C,\C_M}(t-t_0-M\Delta t)\right]
\left[\prod_{i=1}^M G_{\C_i,\C_{i-1}}(\Delta t)\right] \sum_\C G_{\C_0,\C}(t_0) 
p^\mathrm{ss}(\C) \frac1{Z(g,t)}
\label{equ:path_prob}
\end{equation}
where $G_{\C,\C'}(t')=\langle \C|\ee^{\WW(g)t'}|\C'\rangle$  is akin
to a propagator
in the biased ensemble and $p^\mathrm{ss}(\C)$ was defined above 
to be the probability
of observing configuration $\C$ in the (unbiased) steady state of the original
model.  We emphasise that this representation of the path-probability
does not correspond directly to a Markov chain since
$G_{\C,\C'}(t')$ is not a stochastic
matrix [$\sum_\C G_{\C,\C'}\neq 1$ because $\WW(g)$ is
not a stochastic master operator].

Then, within the TTI regime of the biased ensemble, we have from
(\ref{equ:tti_proj}) that
\begin{equation}
P[\pi,g] \approx u(\C_M)
\left[\prod_{i=1}^M G_{\C_i,\C_{i-1}}(\Delta t)\right] v(\C_0)
\frac1{\ee^{-\psi(g)M\Delta t}}
\end{equation}
where the approximate equality is exact in the TTI regime, with the exponentially
small corrections given in (\ref{equ:tti_proj}).
Finally, we define the propagator of the auxiliary model,
and using the definition of $\WW^\mathrm{aux}$, we have
\begin{equation}
G^\mathrm{aux}_{\C,\C'}(t') \equiv \langle \C|\ee^{\WW^\mathrm{aux}(g)t'}|\C'\rangle
= u(\C) G_{\C,\C'}(t') \frac{1}{u(\C')} \ee^{\psi(g) t'}
\end{equation}
This represents a stochastic propagator,
in the sense that
$\sum_\C G^\mathrm{aux}_{\C,\C'}(t)=1$.  Thus, the path probability in the
biased ensemble is
\begin{equation}
P[\pi,g] \approx 
\left[\prod_{i=1}^M G^\mathrm{aux}_{\C_i,\C_{i-1}}(\Delta t)\right] p^\mathrm{TTI}(\C_0)
\label{equ:aux_path}
\end{equation}
where the approximate equality is exact in the TTI regime, as above.
We recognise the right hand side of (\ref{equ:aux_path})
as the path probability for the (stochastic)
auxiliary model in its steady state.

Thus, we have shown that the steady state associated with
the biased ensemble of trajectories can be interpreted as
the steady state of an auxiliary stochastic model that is
Markov and TTI.  Transition rates that are non-zero in the original
model are also non-zero in the auxiliary model, and vice versa.
(For example, if the original model has only single spin flip moves then
so does the auxiliary model, and if some moves are forbidden by kinetic
constraints in the original model~\cite{kcm,s-glass} 
then these constraints are preserved in the auxiliary model.)
However, we note that the auxiliary model may be unphysical
in the sense that the dynamical rules are non-local.  For example,
in stochastic spin models the rate for flipping a given spin
typically depends only on the states of that spin and of a few
spins in its neighbourhood.  Even if the original
model $\WW$ is constructed in this way, the auxiliary
model $\WW^\mathrm{aux}$ typically does not share this feature.
In general, it is not clear if such
non-local interactions render these biased ensembles unphysical,
or if they might arise from non-trivial fluctuation
forces in non-equilibrium states, as proposed by Evans~\cite{Mike-Evans}.

We end this section with a note about causality: if we consider
an unbiased ensemble of trajectories that begins at equilibrium
but is perturbed (by a change in its transition rates) at some
time $t_\mathrm{w}$, then the response to the perturbation is
zero for all times $t'<t_\mathrm{w}$.  However, if we then
use these perturbed trajectories to generate a biased
ensemble as in (\ref{equ:s-ens}), then one will generically observe a response to
the perturbation for times $t'<t_\mathrm{w}$.  This
is the violation of causality that was mentioned in the introduction.  
Clearly then, a
perturbation to the original transition rates at time $t_\mathrm{w}$
cannot be taken into account through a perturbation at time
$t_\mathrm{w}$ in the auxiliary model. 

\subsection{Time-reversal invariance and a variational result}

So far, we have considered a general stochastic model.  We
now focus on ensembles of trajectories that are time-reversal
invariant.  For the unbiased ensemble of trajectories, this
condition is met for stochastic models obeying detailed balance:
$W(\C\to\C') \ee^{-\beta E(\C)} = W(\C'\to\C) \ee^{-\beta E(\C')}$
where $E(\C)$ is the energy of configuration~$\C$, and $\beta$ is the
inverse temperature, as above.
In this case, biased ensembles of
trajectories defined as in (\ref{equ:s-ens}) are also time-reversal
invariant, and the operator $\WW(g)$ may be symmetrised
by a similarity transformation:
\begin{equation}
\HH(g) = \ee^{\beta\hat E/2} \WW(g) \ee^{-\beta\hat E/2}
\label{equ:Wsym}
\end{equation}
where $\hat E$ is the (diagonal) energy operator.
Such a symmetrisation is possible because the bias introduced
in (\ref{equ:s-ens}) is itself time-reversal symmetric.  
One may also consider biasing ensembles according
to observables $\Acal_t$ defined as in (\ref{equ:obsA}). In this
case, if the unbiased model obeys detailed balance then
the biased ensemble is time-reversal symmetric only
if the bias is also time-reversal symmetric: $\alpha(\C,\C')=\alpha(\C',\C)$.
An example is the case where $\Acal_t$ is simply the number of
configuration changes in the trajectory, with $\alpha(\C,\C')=1$ for
all pairs of distinct configurations.  However, if $\Acal_t$ is a current or a flux
then time-reversal symmetry is typically broken, and $\WW(g)$
may not be symmetrised.

We also note that if the rates $W(\C\to\C')$ obey 
detailed balance then so do the rates of
the auxiliary model.  
Since $\HH(g)$ is symmetric, 
the left and right eigenvectors of $\WW(g)$ are related
through $v(\C)=u(\C)\ee^{-\beta E(\C)}$. Expressing $\HH(g)$ in terms of 
$\WW^\mathrm{aux}$ and multiplying it from both sides first by 
$\ee^{-\beta\hat E/2}$ and then by $\hat{U}$ also shows that
$\WW^\mathrm{aux}\hat{U}\ee^{-\beta\hat E}\hat{U}$ is symmetric, i.e.
\begin{equation}
W^\mathrm{aux}(\C'\to\C) u(\C') v(\C')
= W^\mathrm{aux}(\C\to\C') u(\C) v(\C)
\end{equation}
Thus, the auxiliary rates $W^\mathrm{aux}(\C\to\C')$ obey
detailed balance with respect to the steady state distribution
$p^\mathrm{TTI}(\C)$ defined above, as claimed.  We 
therefore define an auxiliary energy function $ E^\mathrm{aux}$
through
\begin{equation}
\ee^{-\beta E^\mathrm{aux}(\C)}=u(\C) v(\C) = \ee^{-\beta E(\C)} u(\C)^2
\end{equation}
so that the difference in the energy of a configuration
between unbiased and auxiliary models is $E^\mathrm{aux}(\C)-E(\C)=(-2/\beta)\ln u(\C)$.

We emphasise that the transition rates in the auxiliary model
can be obtained from the rates of the original model using only
the auxiliary energy function as additional information.  
Further, since the operator $\HH(g)$ is symmetric, one may
obtain the dynamical free energy through a variational principle
\begin{equation}
-\psi(g) = \max_{|\phi\rangle} \frac{\langle \phi|\HH(g)|\phi\rangle}
 {\langle \phi|\phi\rangle}
\end{equation}
with equality when $|\phi\rangle=\sum_\C \sqrt{u(\C) v(\C)} |\C\rangle$.
This may be interpreted as an extremisation over variational
estimates for the steady state distribution through
$|\phi\rangle=\sum_\C \sqrt{p^\mathrm{var}(\C)}|\C\rangle$,
with equality when  $p^\mathrm{var}(\C)=p^\mathrm{TTI}(\C)$.
Equivalently, one may consider an extremisation over energy functions
$|\phi\rangle=\sum_\C \ee^{-\beta E^\mathrm{var}(\C)/2}|\C\rangle$
with equality when $E^\mathrm{var}(\C)=E^\mathrm{aux}(\C)$.
However, as noted above, models with short-ranged interactions
have energy functions consisting of sums over local contributions,
but there is no reason to suppose that $E^\mathrm{aux}(\C)$ can
be written in this way.  A specific example will be given
in the next section.  

\section{$1d$ Glauber-Ising chain, and link with quantum-Ising chain}
\label{sec:ising}

To illustrate the general features described above, we
consider the $1d$ Glauber-Ising chain.  We take a periodic chain
of $N$ Ising spins, $s_i=\pm1$ with an energy function
$E=-\hf\sum_i s_i s_{i+1}$.  Spins flip with Glauber rates, respecting
detailed balance: the
rate for flipping spin $i$ is
$W_i=[1+\exp(h_i s_i)]^{-1}$ where $h_i=\beta(s_{i-1}+s_{i+1})$.
It will also be useful to define domain wall variables 
$n_{i+\hf}=\hf(1-s_i s_{i+1})$, so that 
$n_{i+\hf}=1$ if there is a domain wall between sites $i$ and
$i+1$ while $n_{i+\hf}=0$ otherwise.  Periodic boundaries ensure that
the total number of domain walls in the system is even, while
we note that every configuration of the domain walls corresponds
to two different configurations of the spins, related by flipping
all of the spins at once.  In terms of the domain wall variables,
the energy is simply $E=\sum_i (n_{i+\hf}-\hf)$.

\subsection{Large deviations of the energy and a dynamical phase transition}

Starting from the Glauber-Ising chain, we
define a biased ensemble of trajectories as in (\ref{equ:s-ens}), taking the
observable $B(t)$ to be the energy $E(t)$.  We write the
master equation in terms of domain-wall variables,
respresenting states by associating a spin-half degree of freedom
with each bond on the chain.  Thus, a domain wall on bond
$i+\hf$ corresponds to an up spin, while bonds with $n_{i+\hf}=0$
correspond to down spins.
Then, the operator (\ref{equ:WB}) for the biased ensemble 
has a representation in terms of
Pauli matrices $\sig^x_{i+\hf},\sig^y_{i+\hf},\sig^z_{i+\hf}$:
\begin{eqnarray}
\mathbb{W}(g) & = 
\hf \sum_i \Big[ &
    \sig^+_{i-\hf} \sig^-_{i+\hf} + \sig^+_{i-\hf} \sig^-_{i+\hf}
 + \gamma \sig^-_{i-\hf} \sig^-_{i+\hf} + 
   \lambda \sig^+_{i-\hf} \sig^+_{i+\hf} 
\nonumber \\ & &
  + (\lambda-1-g) \sig^z_{i+\hf} -1 \Big]
\label{equ:WE_ising}
\end{eqnarray}
with $\sig^\pm_{i+\hf}=\hf(\sig_{i+\hf}^x\pm\ii\sig_{i+\hf}^y)$ as usual,
$\gamma=2/(1+\ee^{-2\beta})$ 
and $\lambda=2-\gamma=2/(1+\ee^{2\beta})$.
We note that large deviations of the energy in
a similar model with an antiferromagnetic interaction
potential may be obtained by
taking $\beta<0$ and flipping the sign of $g$ (since 
$E=+\hf\sum_i s_i s_{i+1}$ in that case).
Symmetrising $\WW(g)$ as in (\ref{equ:Wsym}), one may
diagonalise $\mathbb{H}(g)$ using a Jordan-Wigner transformation
(as, for example, in section 4.2 of Ref.~\citen{SachdevQPT}, leading to 
\begin{equation}
\HH(g) = \sum_{k} \left[ \Omega_k (c_k c^\dag_k- c^\dag_k c_k) - \hf \right]
\end{equation}
where $c^\dag$ and $c$ are fermionic creation and annihilation
operators, labelled by a wave vector $k=2m\pi/N$, the sum runs
over integer $m$ in the first Brillouin zone, $-\frac {N}2 < m \leq \frac N2$,
and
\begin{equation}
\Omega_k = \sqrt{(g-\lambda +1- \cos k)^2 + \lambda\gamma \sin^2 k}.
\label{equ:Omega_k}
\end{equation}
We have $\Omega_k\geq0$ so that the largest eigenvalue of $\HH$
is $-\psi(g)= \sum_k (\Omega_k-\hf)$.  The fermion vacuum state has such
an eigenvalue, and we identify the difference between the largest and
next-largest eigenvalues as $\Delta\omega=\min_k (2\Omega_k)$.  Finally,
taking the limit of large system size,
\begin{equation}
\psi(g) = \frac{N}2 \int_{-\pi}^\pi \frac{\mathrm{d}k}{2\pi} (1-2\Omega_k)
\end{equation}

\begin{figure}
\begin{center}
\includegraphics[width=6cm]{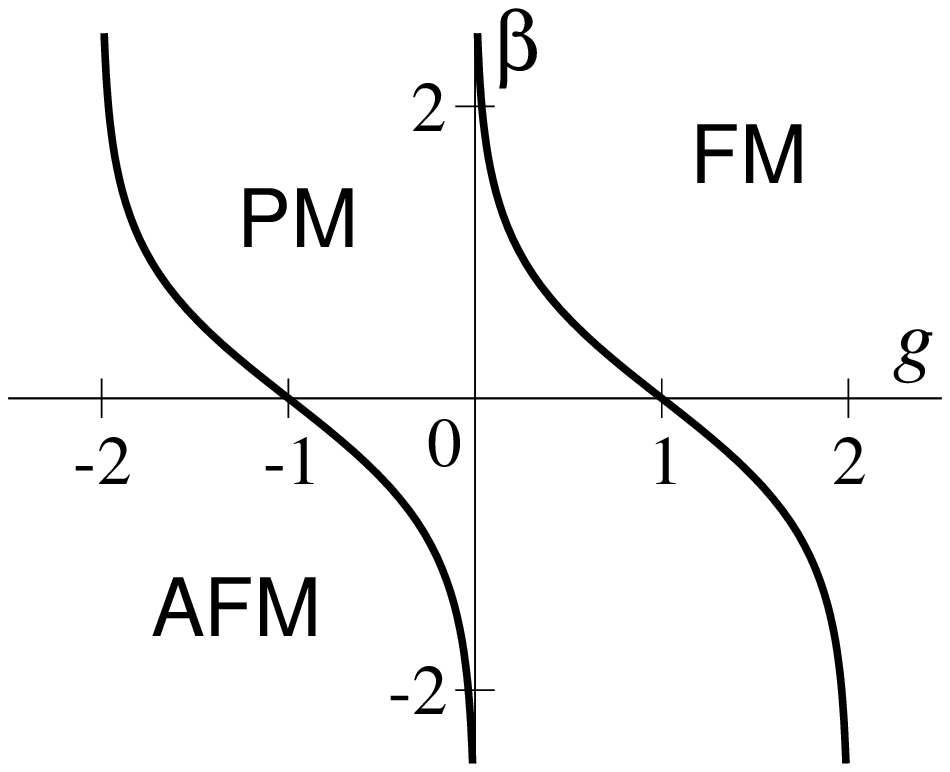}
\includegraphics[width=6cm]{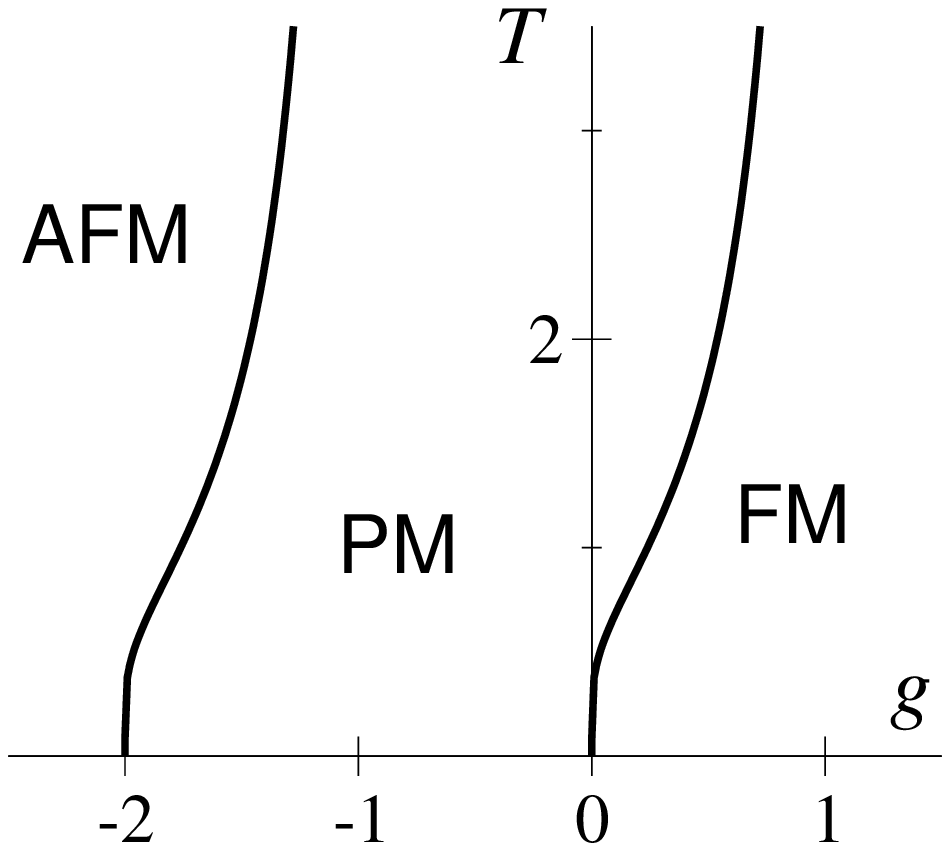}
\end{center}
\caption{Dynamical phase diagram for the 1d Glauber-Ising chain,
as a function of biasing parameter $g$. Solid lines
correspond to second-order (critical) phase transitions.
(Left) We show the
behaviour as a function of inverse temperature $\beta$: the
regime with $\beta<0$ corresponds to an antiferromagnetic
chain, while $\beta=0$ is the chain at infinite temperature.
The equilibrium steady state corresponds to $g=0$ and
is always paramagnetic (PM).  Large positive $g$ drives
the system into a ferromagnetic (FM) phase while
large negative $g$ leads to an antiferromagnetic (AFM) phase.
On the right, we show the same figure as a function of temperature,
to emphasise that both transitions persist all the way to
$T=0$.  However, the nature of the critical points changes
at $T=0$ since the system is no longer time-reversal symmetric.}
\label{fig:phase}
\end{figure}

As well as the dynamical free energy $\psi(g)$, we emphasise that we have
obtained the full spectrum of the operator $\WW(g)$.   A particularly important
case for large system sizes is obtained when $\min_k \Omega_k=0$, so that there are eigenvalues
of $\WW(g)$ that are arbitrarily close to its maximal eigenvalue $-\psi(g)$. 
The essential point here is that $\Omega_k>0$ except in certain
special cases: (i) when $k=0$ and $g=\lambda$, (ii) when $k=\pi$ and $g=-\gamma$ 
(recall $\gamma+\lambda=2$) and (iii) when $\lambda=0$ is equal to 
zero and $\cos k = 1+g$ (or $\gamma=0$ and $\cos k = g-1$). 
 Case~(iii) corresponds to the zero-temperature
dynamics of the Ising model, which we do not consider in this work.  
However, we will consider the cases $g=\gamma>0$ and  $g=-\lambda<0$,
for which the system becomes critical.
It may be verified
that the second derivative of $\psi(g)$ diverges at these
critical points
(note that we have taken this derivative after the limit of large $N$).
This divergence signals a (continuous) phase transition in the space 
of trajectories. In particular, (\ref{equ:tti_proj}) breaks down, and the 
existence of the TTI regime described in Section~\ref{sec:tti} is no longer assured
(this point will be discussed further in Sec.~\ref{sec:ising_phys} below).

Now, the $1d$ Ising model clearly
has no phase transition at any finite temperature, so the equilibrium
ensembles of trajectories with $g=0$ are those of ergodic paramagnets.
However, for $g>\lambda$, the biased ensembles of
trajectories are dominated by ferromagnetic configurations:
the system acquires
long-ranged order in space, and it also breaks ergodicity, exhibiting
long-ranged order in time.   A similar
effect occurs for $g<-\gamma$, for which the system
enters an antiferromagnetic state.  The `dynamical phase diagram' 
for this system is shown in Fig.~\ref{fig:phase}.  Similar critical
phase transitions in biased ensembles have been found in 
higher-dimensional ferromagnets, for temperatures above their 
critical points~\cite{p2-paper,fred-jsp}.  However, to
explain how the physical conclusions described in this section
can be drawn directly
from the form of $W(g)$, we now show
that biased ensembles of trajectories in the $1d$ Ising chain
have already been studied extensively 
in the context of quantum phase transitions~\cite{SachdevQPT}.
This allows us to identify the universality class of the 
(continuous) phase transitions at $g=\gamma$ and $g=-\lambda$: as long as we take
$T>0$ in the stochastic model, this is the universality class
of the (classical) $2d$ Ising model.

\subsection{Link with quantum Ising model in a transverse field
and hence with the $2d$ Ising model}

We begin by taking $\beta=0$ in
the original Ising chain, corresponding to infinite temperature,
so that all single spin-flip transitions take place with rate $\hf$.
In this case, we have, using the superscript ($n$) to emphasize that we are
in the domain wall basis,
\begin{equation}
\HH^{(n)}(g) = \WW^{(n)}(g) = \hf \sum_i \left[ 
\sig^x_{i+\hf} \sig^x_{i-\hf} -1 - g \sig^z_{i+\hf} \right]
\label{equ:qu-ising-n}
\end{equation}
One may also work in the spin basis, using a spin-half degree
of freedom for each of the original Ising spins:
for $\beta=0$, one has a master operator
\begin{equation}
\HH^{(s)}(g) = \WW^{(s)}(g) = \hf \sum_i \left[
\sig^x_i-1 + g \sig^z_i \sig^z_{i+1} \right]
\label{equ:qu-ising}
\end{equation}
These operators are familiar from studies of
quantum spin chains.
To be precise, $[-\HH^{(s)}(g)]$ is the Hamiltonian for
an Ising ferromagnet in transverse field,
a canonical model for quantum phase transitions (QPTs)~\cite{SachdevQPT}.
In $[-\HH^{(s)}(g)]$, one identifies $g$ as 
an Ising coupling between spins, while the term proportional
to $\sigma^x$ is a `transverse field' that frustrates ferromagnetic
ordering.  As $g$ is tuned through $g=1$, the ground
state energy $\psi(g)$ has a singularity, and the system
acquires long-ranged ferromagnetic order.  The point $g=1$ is critical
in that it exhibits a diverging correlation length~\cite{SachdevQPT}. Representing
the partition function for the quantum system
as an imaginary time path integral, one sums
over an ensemble of periodic trajectories of the Ising chain.  
The temporal extent
of these trajectories is given by $\beta_Q \hbar$ where $\beta_Q$
is the inverse temperature of the quantum system and $\hbar$ is 
Planck's constant divided by $2\pi$.  In the limit of 
large $\beta_Q$ (small temperature in the quantum system), one may
consider paths of length $M\Delta t\ll \beta_Q \hbar$ and
show that the (real-valued) weight associated with a path $\pi$ in the quantum 
path integral is proportional to the path probability in the biased ensemble of the Glauber-Ising 
chain, defined as in (\ref{equ:path_prob}).  [The constant of proportionality
is simply the partition function $Z(g,t)$.]

Further, the dynamical free energy $\psi(g)$ can also be identified as the
thermodynamic free energy of a two-dimensional Ising model on a square
lattice.  To arrive at this standard result~\cite{SachdevQPT}, 
consider a $2d$ Ising model with energy 
$E=-\sum_{ij} (J s_{ij} s_{i+1,j} + K s_{ij} s_{i,j+1})$ where the
Ising spins $s_{ij}$ now
carry two indices, indicating their co-ordinates on a square lattice.  Starting
from a system with $J=K$, one may take the lattice
spacing in the $y$-direction to zero, with the couplings $J$ and $K$
being adjusted to keep a constant free energy density.  The result
is a model with a continuous $y$ co-ordinate: one then
identifies this co-ordinate
with the time $t$ in the quantum model or the Glauber-Ising chain~\cite{SachdevQPT}.
Comparing (\ref{equ:qu-ising-n}) and (\ref{equ:qu-ising}), there is clearly a duality between
models with biasing parameters $g$ and $1/g$:
the mapping to the square lattice Ising model allows us to identify
this as the Kramers-Wannier duality of the $2d$ Ising model.  We note in passing that
while the free-fermion solution for the large deviation function $\psi(g)$ is possible
only for the $1d$ Glauber-Ising model, the mapping from a $d$-dimensional
Glauber-Ising model to a $d$-dimensional quantum spin model applies in all dimensions,
and the critical properties of these $d$-dimensional models are the same as those
of a $d+1$-dimensional classical Ising model~\cite{SachdevQPT}.
(However, the mapping in $d>1$ takes place at
the level of an effective field theory, so it applies only to universal quantities.)

If one now works at finite temperature for the Glauber-Ising chain,
one obtains $\HH^{(n)}(g)$ by symmetrising
the operator $\WW^{(n)}(g)$ given in (\ref{equ:WE_ising}), and one may
also write down the operators $\WW^{(s)}(g)$ and $\HH^{(s)}(g)$.  
Compared to the infinite temperature case,
the symmetrised operators contain extra terms, but these are all
irrelevant in the renormalisation group 
sense.  It follows that for $T>0$, the dynamical phase transitions 
shown in Fig.~\ref{fig:phase} are in the universality
class of the quantum-Ising chain or, equivalently, the $2d$ Ising model.

Finally, we note that these mappings break down in the special case
where $T=0$ in the Glauber-Ising chain: this model
corresponds to a reaction-diffusion system $A+A\to0$.  This
is a non-equilibrium critical system in the sense that the decay
of finite-density initial conditions 
towards the ground state occurs in a power-law fashion and 
involves a dynamical scaling length that grows as a power
of the time.  In this case, detailed balance is not obeyed and
the operator $\WW(g)$ may not be symmetrised.  However, the model
may still be solved by free fermions and the phase diagrams show that
paramagnetic, ferromagnetic and antiferromagnetic phases may all be
observed at zero temperature.
We postpone a discussion of
these biased enembles to a later study.

\subsection{Physical interpretation of the biased ensemble
in the Glauber-Ising chain}
\label{sec:ising_phys}

It follows from the above analysis that if we take a large
Glauber-Ising chain and select long trajectories with a small value 
of the time-integrated energy, this ensemble is dominated by
ferromagnetic trajectories that spontaneously break
the $Z_2$ symmetry of the Ising chain.  To understand the effects
of this symmetry breaking, It is instructive to 
consider the master operator in the basis of the original Ising spins.
The paramagnetic
phase corresponds to a non-degenerate largest eigenvalue of
$\WW^{(s)}(g)$ and (\ref{equ:Omega_k}) indicates that there is a finite gap
between the largest and
second-largest eigenvalues.  Thus, (\ref{equ:tti_proj}) holds, and the analysis
of Sec.~\ref{sec:tti} follows.  On the other hand, at the critical point, 
there is no gap in the spectrum, and both
 (\ref{equ:tti_proj}) and the TTI regime break down (for large system size $N$).  
 Then, in the ferromagnetic
 phase, spontaneous symmetry breaking means that the largest eigenvalue of
 $\WW^{(s)}(g)$ is now doubly degenerate, but all other eigenvalues are separated
 from these two by a finite gap.  In that case, (\ref{equ:tti_proj}) may be generalised
 into a projection onto the lowest two eigenvectors of $\WW^{(s)}(g)$, and the existence
 of a TTI regime may again be proven.  This illustrates the point that we made
 in Sec.~\ref{sec:tti}, that a non-degenerate largest eigenvalue of $\WW(g)$ is
 sufficient to ensure the existence of a TTI regime, but it is not necessary.

Furthermore, from the analysis of Section~\ref{sec:general}, the ensemble of
trajectories that we have defined here by biasing the time-integrated energy
can also be generated by an auxiliary (Markov)
stochastic model that respects detailed balance with
respect to its steady-state distribution.  Since $1d$ systems
with short-ranged interactions do not permit ferromagnetic
states, it follows that the effective energy function 
$E^\mathrm{aux}$ associated with the auxiliary model must 
contain long-ranged interactions.  In fact, the form
of $E^\mathrm{aux}$ has been discussed quite
extensively in the mathematical physics literature
for the closely-related case of a single-layer 
in a $2d$ Ising model on a square lattice~\cite{Maes-square}.

\section{Conclusion}

In this article, we have brought together several results
for biased ensembles, defined as in (\ref{equ:s-ens}).  Our main interest
concerns the degree to which these ensembles represent 
physically-reasonable dynamics that might be sampled
by some experimental procedure.

For general biased ensembles, we showed
that one may always construct an auxiliary Markov process whose steady
state reproduces the TTI regime of the biased ensembles.
Transition rates in the auxiliary models are modified from their
original values by factors that depend only on a single
left-eigenvector of a master operator $\WW(g)$.  (Of course,
obtaining this eigenvector is likely to be impossible
except in relatively simple exactly-soluble models).  In any case, the 
biased ensembles are Markov, although the rates for local processes
may depend on configurations of the system in far away regions.

For biased ensembles that respect time-reversal symmetry, we showed
that this auxiliary Markov process may always be constructed
in terms of an auxiliary energy function, and we established a 
variational bound on this energy function.  However, it is likely
that this auxiliary energy function typically contains non-local
interactions.  The presence of such interactions is proven for
the Glauber-Ising chain since the model undergoes a phase
transition to a ferromagnetic state. 

The crucial question arising from Refs.~\citen{merolle,s-glass}
is whether the presence
of phase transitions in biased ensembles of trajectories can be used to
explain the properties of the original (unbiased) stochastic model.
In the Ising chain, one may intepret the low temperature behaviour
in terms of patches of the two ferromagnetic phases, while noting
that large enough ferromagnetic domains are always unstable to
thermal fluctuations.  We have shown that biased ensembles
of trajectories can stabilise the ferromagnetic phases,
and reveal the symmetry between them.  In the context of the glass
transition, one might argue that biased ensembles are similarly
effective in revealing states that are eventually unstable to thermal
fluctuations, but nonetheless persist for long enough
to explain the large relaxation times observed in glass-forming
liquids.  It would certainly be very interesting to understand
what terms appear in the auxiliary energy function as glassy
systems break ergodicity, and to study how these interactions
stabilise the amorphous solid phase.

\section*{Acknowledgements}

We thank David Chandler, Mike Evans, Juan Garrahan, Thomas Speck and
Fred van Wijland for helpful discussions. 
We also thank the organisers of the workshop ``Frontiers in non-equilibrium
physics'' at the Yukawa Institute in Kyoto for their support for this work.
RLJ is also grateful for 
financial support from the Franco-British Alliance programme,
managed by the British Council and the French
Foreign Affairs Ministry (MAE).


\begin{thebibliography}{99}

\bibitem{Bod-Der}
T.~Bodineau and B.~Derrida, Phys. Rev. Lett. {\bf 92} (2004), 180601.

\bibitem{Ruelle} D. Ruelle, ``{\it Thermodynamic Formalism}''
   (Addison-Wesley, Reading, 1978); J.-P. Eckmann and D. Ruelle,
   Rev. Mod. Phys. {\bf 57} (1985), 617.

\bibitem{fluct}
G.~Gallavotti and E.~G.~D.~Cohen, J. Stat. Phys. {\bf 80} (1995), 931;
C.~Jarzynski, Phys. Rev. Lett. {\bf 78} (1997), 2690;
J.~Kurchan, J. Phys. A {\bf 31} (1998), 3719;
G.~E.~Crooks, Phys. Rev. E {\bf 61} (2000), 2361.

\bibitem{JL} 
L.~Bertini, A.~De Sole, D.~Gabrielli, G.~Jona-Lasinio
and C.~Landim, J. Stat. Phys. {\bf 135} (2009), 857.

\bibitem{JL-control}
L.~Bertini, A.~De Sole, D.~Gabrielli, G.~Jona-Lasinio
and C.~Landim, J. Stat. Phys. {\bf 116} (2004), 831.

\bibitem{Lebowitz-Spohn}
J.L. Lebowitz and H. Spohn, J. Stat. Phys. {\bf 95} (1999), 333.

\bibitem{Maes-general}
C.~Maes, J. Stat. Phys. {\bf 95} (1999), 367;
C.~Maes and M.~H.~Wieren, Phys. Rev. Lett. {\bf 96} (2006), 240601.

\bibitem{ECM-93}
D.~J.~Evans, E.~G.~D.~Cohen and G.~P.~Morriss, Phys. Rev. Lett. 
{\bf 71} (1993), 2401.

\bibitem{Mike-Evans}
R.~M.~L.~Evans, Phys. Rev. Lett. {\bf 92} (2004), 150601;
A.~Simha, R.~M.~L.~Evans and A.~Baule, Phys. Rev. E {\bf 77} (2008), 031117.

\bibitem{merolle}
M. Merolle, J.P. Garrahan and D. Chandler,
   Proc. Natl. Acad. Sci. USA {\bf 102} (2005), 10837; 
 R.L. Jack, J.P. Garrahan and D. Chandler, 
   J. Chem. Phys. {\bf 125} (2006), 184509.

\bibitem{s-glass}
J.P. Garrahan \emph{et al.},
Phys. Rev. Lett. \textbf{98} (2007), 195702;
L.~O.~Hedges \emph{et al.}, Science {\bf 323} (2009), 1309.

\bibitem{p2-paper}
K.~van Duijvendijk, R.~L.~Jack and F.~van Wijland, 
Phys. Rev. E {\bf81} (2010), 011110. 

\bibitem{rom-paper}
R.~L.~Jack and J.~P.~Garrahan,
Phys. Rev. E {\bf81} (2010), 011111. 

\bibitem{juanpe-fred-jpa}
J.P. Garrahan \emph{et al.}, J. Phys. A {\bf 42} (2009), 075007.

\bibitem{fred-jsp}
V.~Lecomte, C.~Appert-Roland and F.~van Wijland, J. Stat. Phys. {\bf 127} (2007), 51.

\bibitem{Gaveau-Schulman}
B.~Gaveau and L.~S.~Schulman, J. Math. Phys. {\bf 39} (1998), 1517.

\bibitem{absorbing}
H.~Hinrichsen, Adv. Phys. {\bf49} (2000), 815.

\bibitem{kcm}
 F.~Ritort and P.~Sollich, 
  Adv. Phys. {\bf 52} (2003), 219.

\bibitem{SachdevQPT}
S. Sachdev, {\em ``Quantum Phase Transitions''} (Cambridge University Press,
Cambridge, UK, 1999).

\bibitem{Maes-square}
C.~Maes and F.~Redig and A.~van Moffaert, J. Stat. Phys. {\bf96} (1999), 69. 

\end{thebibliography}
\end{document}